\documentclass[aps,showpacs,preprint,superscriptaddress,eqsecnum]{revtex4}

\usepackage{epsfig}
\usepackage{graphicx}
\usepackage{bm}

\begin{document}

\title{Quantum-state transfer on spin-chain channels with random imperfections}

\author{De-Xin Kong}
\affiliation{Department of modern physics, University of Science and
Technology of China, HeFei, 230026,P.R.China}
\author{An Min Wang}
\email{anmwang@ustc.edu.cn} \affiliation{Department of modern
physics, University of Science and Technology of China, HeFei,
230026,P.R.China}


\begin{abstract}
We investigate the quantum-state transfer on spin-chian channels
with random imperfections.Through combining the advantages of two
known schemes, the dual-rail spin-chain channels\cite{parallel} and
the particular ihhomogenous spin-chain channel\cite{endsc}, we
propose a protocol that can avoid the quantum noises introduced by
many unnecessary measurements and can enhance the anti-decoherence
ability. The results show that our protocol is more efficient to
transfer an arbitrary quantum state than the original one. In
particular, we discuss the effects of couplings fluctuations and
imperfect initialization on both of the improved scheme and original
one.
\end{abstract}


\maketitle

\section{Introduction}
The quantum states transfer from one location to another is
necessary in many quantum information processing systems. Therefore,
a number of technologies have been developed to accomplish this
task. One of the most developed systems is optical system, which is
employed to transfers quantum states directly via photons. For
example, photons in cavity\cite{opticav} and in ion
traps\cite{opticion} were used as a variety of information carriers
to transfer quantum states from site to site. And also photons could
be used to create entanglement between two sites for further
teleportation\cite{optictel}. Even though these optical technologies
are very efficient in transferring quantum states through long
distance, employing them to transfer quantum states through short
distance within a quantum computer or quantum information processing
system, may cause some very upsetting interfacing problems (between
optical carrier and solid state matter) for engineering the quantum
information processing system. However, quantum communication among
different parts (the processor and the memory) of a quantum computer
is always necessary\cite{qst}. The following types of quantum
channels are developed to accomplish the task of quantum
communication among different parts of a quantum computer.

In principle, both a series of swap gates and the spin
chain\cite{sc1} can be employed to transfer quantum states. However,
it is impractical to use a series of swap gates as the quantum
channel in reality. In Ref.\cite{sc1}, Sougato Bose suggested that
one can use a wire of spins interacting equally with their nearest
neighbors as quantum channels to transfer quantum states and
entanglement under the free evolution, but this scheme can not
transfer quantum states perfectly, and the fidelity of transferring
states would decrease with the length of a spin chain. However, In
our opinion, the advantages of using the spin chain as quantum
channel make this protocol still be a promising technology. These
advantages are: avoiding interfacing problems and extremely
decreasing external control in the process of transferring states.
Because of the above two features, a series of inhomogenous spin
chains\cite{Mattias, inhomo} with engineering coupling constants
which can transfer quantum states perfectly, were developed.
Although these inhomogenous spin chains can transfer quantum states
perfectly, engineering them in experiment is very difficult.

Several other methods have been developed to improve the transfer
fidelity of spin-chain channels. Exposing a spin chain system to
external modulated magnetic field\cite{magnet} can also
significantly improve the fidelity of transferring quantum states.
In Ref. \cite{twogate}, the authors proposed a protocol that can
optimizing the fidelity by applying a suitable sequence of two-qubit
gates at the receiving end of the chain. This protocol is
interesting because the two-qubit gates are simple and can be
realized by a simple switchable interaction, but the perfect state
transfer requires infinite number of operations. In addition,
parallel spin-chain channels\cite{parallel} can be used to perform
conclusive and arbitrarily perfect quantum-state transfer. Their
protocol used only a parallel spin-chain channel composed by two
uniform spin chains without interacting with each other, and some
local operations and measurements. This scheme is natural and
simple, but the high probability of success to perfectly transfer a
quantum state definitely implies large number of measurements and
long transferring time, which renders the scheme impractical in
experiment, especially for long spin chains---another factor of
affecting the efficiency of the protocol. For instance, for $N=150$
(the number of spins in one of the spin chains), one needs to
perform near \textsl{30} measurements so that the probability of
success can be achieved \textsl{90\%}.

Obviously, because of the limitation of measurement technology, the
measurement can generally lead to the departure of resulted states.
Here, we call the departure introduced by measurements as quantum
noise. In this paper, our aim is to find a way to avoid the quantum
noise perhaps introduced by many measurements, but without
decreasing the efficiency of quantum-state transfer and losing the
scheme's natural and simple feature. Inspired by the
work\cite{endsc} of Antoni W\'{o}jcik, et al, who has shown that one
can greatly improve the transfer fidelity through the spin-chain
channel by modulating the parameter $a$ (the coupling strength of
both ends of a spin chain with the other uniform part of the spin
chain). Our proposal is to substitute the homogenous spin chains of
the original protocol with the spin chains in Ref. \cite{endsc}, and
to add some requirements to the original scheme. We also have
studied the effects of imperfections caused by couplings
fluctuations and imperfect initialization. Another important
advantage of this new scheme is that it maintains natural and simple
feature.

This paper is organized as follows. In Sec. \ref{sec2}, we briefly
review the work of \cite{parallel}. In Sec. \ref{sec3}, the improved
scheme is studied and advantages of the improved scheme are
discussed. In Sec. \ref{sec4}, we mainly study the effects of
couplings fluctuations and imperfect initialization to the
efficiency of quantum-state transfer and show the improved protocol
is more robust to these kinds of imperfections than the original
one. Conclusions and discussions are made in Sec. \ref{sec5}.

\section{Review of the previous work\cite{parallel}}\label{sec2}
The authors of \cite{parallel} proposed a protocol for perfect but
undetermined quantum communication, which was called
\emph{conclusive} transfer, by encoding an arbitrary quantum states
into a dural-rail quantum channel composed by two parallel uniform
spin chains and by only performing some unitary operations and
measurements on the destination of the channel. The schematic figure
of the protocol is shown in Fig. 1\cite{parallel}.
\begin{figure}
\includegraphics[0pt,0pt][397pt,164pt]{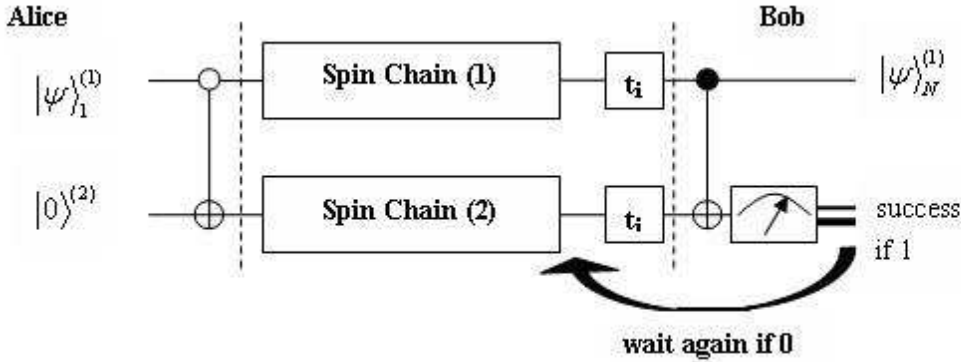} \caption{Quantum
circuit of \emph{conclusive} transfer. The first gate at Alice's
qubits represents a NOT gate to the second qubit controlled by the
first qubit being zero.}
\end{figure}
The Hamiltonian of the suggested system is
\begin{equation}\label{eq:1}
H=H^{(1)}{\otimes}I^{(2)}+I^{(1)}{\otimes}H^{(2)},
\end{equation}
where the superscripts (1) and (2) represent the spin chain (1) and
(2), respectively, and $H^{(1)}$ and $H^{(2)}$ are the identical
Hamiltonians of Heisenberg spin-$\frac{1}{2}$-chain with equal
nearest-neighbor couplings of the spin chain (1) and the spin chain
(2)
\begin{equation}\label{eq:2}
H^{(1)}=H^{(2)}=H_U=-J\sum_{n=1}^N{\bm{\sigma}_n\cdot\bm{\sigma}_{n+1}},
\end{equation}
respectively. The scheme for conclusive transfer is:
\begin{enumerate}
\item \emph{Initialization of the system}. each chain is cooled down
to its ground state, by ground state of the Heisenberg ferromagnetic
chain we mean that all qubits of the chain are downward representing
by $|\mathbf{0}\rangle\equiv|0_1{\ldots}0_N\rangle$.
\item \emph{Encoding the quantum state}. the first qubit of spin
chain (1) is prepared to be
$$|\psi\rangle^{(1)}\equiv\alpha|\mathbf{0}\rangle+\beta|\mathbf{1}\rangle,$$
where $|\bm{n}\rangle$ means that \textsl{n}th qubit is spin-up
while all other qubits of a spin chain are spin-down. The first
qubit of spin chain (2) is in state
$|\psi\rangle^{(2)}=|\mathbf{0}\rangle$. Then Alice applies a NOT
gate to the first qubit of the spin chain (2) controlled by the
first qubit of the spin chain (1) being zero. The state of the
system after the encoding process becomes
$$|\psi(0)\rangle=\alpha|\mathbf{0}\rangle^{(1)}\otimes|\mathbf{1}\rangle^{(2)}+\beta|\mathbf{1}\rangle^{(1)}\otimes|\mathbf{0}\rangle^{(2)}.$$\label{eq:in-code}
\item \emph{Free evolution and measurements}. The system encoded
with the quantum state evolves freely under the Hamiltonian of
(\ref{eq:1}). After the time ${\tau}_1$, the state can be written as
$$|\psi(\tau_1)\rangle=\sum^N_{n=1}{f_{n,1}(\tau_1)|s(n)\rangle},$$\label{eq:evolv}
where
$|s(n)\rangle=\alpha|\mathbf{0}\rangle^{(1)}\otimes|\bm{n}\rangle^{(2)}+\beta|\bm{n}\rangle^{(1)}\otimes|\mathbf{0}\rangle^{(2)}$
and $f_{n,1}\equiv\langle\bm{n}|e^{-iH_Ut}|\mathbf{1}\rangle$. Then
Bob can decode the qubit by applying a C-NOT gate on the
\textsl{N}th site of the dual-rail quantum channels controlled by
the \textsl{N}th qubit of the spin chain (1). The state thereafter
will be
\begin{equation}\label{eq:cnot}
|\phi(\tau_1)\rangle=\sum^{N-1}_{n=1}{f_{n,1}(\tau_1)|s(n)\rangle+f_{N,1}(\tau_1)|\mathbf{\psi}_N^{(1)}\rangle\otimes|\bm{N}\rangle^{(2)}},
\end{equation}
where
$|\mathbf{\psi}_N^{(1)}\rangle\equiv\alpha|\mathbf{0}\rangle^{(1)}+\beta|\bm{N}\rangle^{(1)}$.
The authors' point is that Bob can justify wether the quantum state
$|\psi(0)\rangle$ encoded in the spin chain (1) has been transferred
to the \textsl{N}th site qubit of the spin chain (1) by applying a
measurement on the \textsl{N}th site qubit of the spin chain (2), if
the result of the measurement is ``1", Bob would know that the state
$|\psi(0)\rangle$ has been successfully transferred and then
terminate the transferring process, if the result is ``0", Bob can
conclude that the state $|\psi(0)\rangle$ has not yet been
transferred to the \textsl{N}th site. The result ``0" does not imply
a failure and end of the process of transfer, because, according the
authors' thought, Bob can perform another similar operations on the
spin chain (2) after some time $\tau_2$, since the result ``0" of
the first measurement did not provide any information about the
state $|\psi(0)\rangle$, and therefore, $|\psi(0)\rangle$ is still
residing in the spin chain (1). If the outcome is ``0" again, Bob
can repeat the procedure again and again until the outcome ``1" is
attained. The probability of success with only one measurement is
$|f_{N,1}(\tau_1)|^2$. It has been proved\cite{parallel} that the
probability of success for perfect state transfer increases with the
number of measurements for the uniform Heisenberg spin chain, but
can never achieve one.
\end{enumerate}

The protocol is interesting, for perfect state transfer can also be
realized by using uniform spin chains which can not perfectly
transfer quantum states when $N\ge4$ \cite{Mattias}. However, to
achieve reasonable probability of success for perfect states
transfer, this protocol calls for a large number of measurements
which also related to the length of the channel, therefore the
increase of the length of the channel and of the probability of
success would cause the number of measurements and C-NOT operations
to increase rapidly. As we all know, the external operations and
measurements would definitely cause quantum noises and decoherences,
therefore any protocol involved with so many measurements and
external operations has little significance in reality. Furthermore,
the large number of measurements and operations would take large
amount of time, which has no benefits to complete quantum
information task.

To avoid these disadvantages of the original dual-rail protocol, we
proposed an improved scheme. We will discuss the improved protocol
in details in subsequent sections.
\section{The improved protocol}\label{sec3}
One way to avoid the quantum noise introduced by unnecessary
operations is to increase the probability of success of the first
several measurements, so that the number of operations can decrease
significantly. According to the discussion above, after only one
measurement, the probability of success for perfect transfer is
directly determined by $|f_{N,1}|$, so the efficiency of the
original protocol can be improved by increasing $|f_{N,1}|$. A
special and simple spin-$\frac{1}{2}$ chain of Heisenberg XY model
was proposed and studied in Ref. \cite{endsc}, in which the authors
showed that one can significantly increase $|f_{N,1}|$ by adjusting
the parameter $a$. The Hamiltonian of the spin chain studied in
\cite{endsc} is
\begin{equation}\label{eq:inh-H}
H_{\mathrm{inh}}=\frac{a}{2}(\sigma_1^x\sigma_2^x+\sigma_1^y\sigma_2^y+\sigma_{N-1}^x\sigma_N^x+\sigma_{N-1}^y\sigma_N^y)+\frac{J}{2}\sum^{N-2}_{n=2}{(\sigma_n^x\sigma_{n+1}^x+\sigma_n^y\sigma_{n+1}^y)}+\sum_{n=1}^N{\omega\sigma_n^z},
\end{equation}
where $a$ and $J$ are coupling strength and $\omega$ is Larmor spin
frequency of every site, and $\sigma^i$s (i=x,y,z) are Pauli
matrices. Since $[\sigma_{tot}, H_{\mathrm{inh}}]=0$, where
$\sigma_{tot}=\sum_{n=1}^N{\sigma^z},$ the subspace with single
excitation is invariant and the term
$\sum_{n=1}^N{\omega\sigma_n^z}$ just contributes global phase in
the evolution of the system. For simplicity, we set $J=1$ and
$\omega=0$. Since we find that the transferring fidelity can not be
improved when $a>J$, we just set $0<a<1$. In the subspace of single
excitation, the eigenvalue\cite{endsc} of $H_{\mathrm{inh}}$ is
$\Lambda_k=2\cos(k)$, where \textsl{k} is a solution of either of
the two following equations $(\mu=\pm{1})$
$${\mu}\cot(k)\cot^{\mu}\left( \frac{N-1}{2}k \right)=\frac{a^2}{2-a^2}.$$
The eigenvector $|\upsilon^k\rangle$ corresponding to the eigenvalue
$2\cos(k)$ has the following components:
\begin{eqnarray}
\upsilon^k_1&=&\frac{a}{c}\sin(k),\nonumber\\
\upsilon^k_i&=&\frac{1}{c}\left\{ \sin[(i+1)k]+(1-a^2)\sin[(i-1)k] \right\}\quad 1<i<N,\nonumber\\
\upsilon^k_N&=&\mu\frac{a}{c}\sin(k),\nonumber
\end{eqnarray}
where the normalization factor \textsl{c} reads
$$c^2=(N-1)\left(2(1-a^2)\cos^2(k)+\frac{a^4}{2} \right)+2a^2-a^4.$$
To increase the value of $|f_{N,1}|$, we substitute the uniform spin
chains with the spin chains in Ref.\cite{endsc}, thus the expression
of new $f_{N,1}$ is
$\langle\bm{N}|e^{-iH_{\mathrm{inh}}t}|\mathbf{1}\rangle$, which we
denote as ${f'}_{N,1}$.

\begin{figure}
\includegraphics{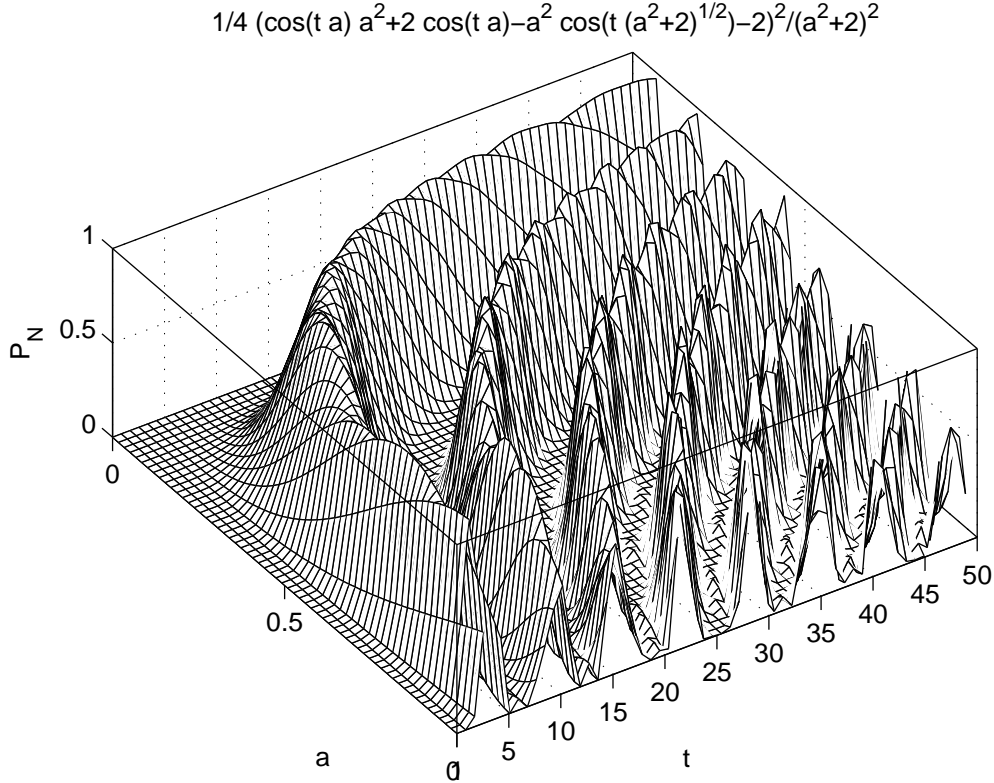}
\caption{\label{fig-2}$P_N$ as a function of parameters $a$ and $t$
for $N=5$.}
\end{figure}

Even though the form of analytical solutions of $H_{\mathrm{inh}}$
was given in Ref. \cite{endsc}, it is really hard to analytically
write out the solutions for large $N$. To show how the probability
of $N$th site qubit in spin-up state,
$P_N=|{f'}_{N,1}|^2=|\langle\bm{N}|e^{-iH_{\mathrm{inh}}t}|\mathbf{1}\rangle|^2$,
is determined by parameters of $a$ and $t$, we set $N=5$, for which
the analytical expression of $P_N$ can be easily handled and
analyzed.
$$P_{N=5}=\frac{1}{4}\frac{\left( \cos(at)a^2+2\cos(at)-a^2\cos\left( t(a^2+2)^{\frac{1}{2}}\right)-2 \right)^2}{(a^2+2)^2}.$$\label{eq:P_n}
In the function of $P_N$, the terms ``$(\cos(at)a^2+2\cos(at)$" and
``$a^2\cos(t(a^2+2)^{\frac{1}{2}})$" will compete with each other,
which causes that the value of $P_N$ would oscillate with the change
of values of $a$ and $t$. In Fig.\ref{fig-2}, we can see the
oscillations. In Ref. \cite{endsc}, the authors showed that this
kind of oscillations always exist for other values of $N$, so the
value of $P_N$ can be improved by adjusting the parameter $a$. The
reason \cite{endsc} for the oscillations is that the change of the
parameter $a$ will vary the distribution of the eigenvectors,
consequently, can influence the process of state transfer.

With this in mind, our changes on the original protocol
\cite{parallel} will state as follows:
\begin{enumerate}
\item \emph{Substitution}. Substitute the uniform spin chains in
original protocol with the spin chains in Ref.\cite{endsc}. Then the
dural-rail quantum channel is composed by two identical inhomogenous
spin chains in Ref.\cite{endsc} uncoupled with each other.
\item \emph{Adjusting the parameter $a$}. To make the improved quantum channel
be the most efficient, one should choose the specific values of $a$,
which relate with the length of the channel.
\item \emph{Only two measurements allowed}. To decrease or limit the quantum noises introduced by
measurements, only two measurements are allowed. We will show that
two measurements are sufficient to make the probability of success
for perfect transfer to achieve more than \textsl{90\%}, and that
the negative effects of extra measurements always outweigh the
benefits of them, so if after the result of the second measurement
still shows failure of the state transfer, one should cease the
process of transferring and initialize the system to restart again.
\end{enumerate}

\begin{figure}
\includegraphics{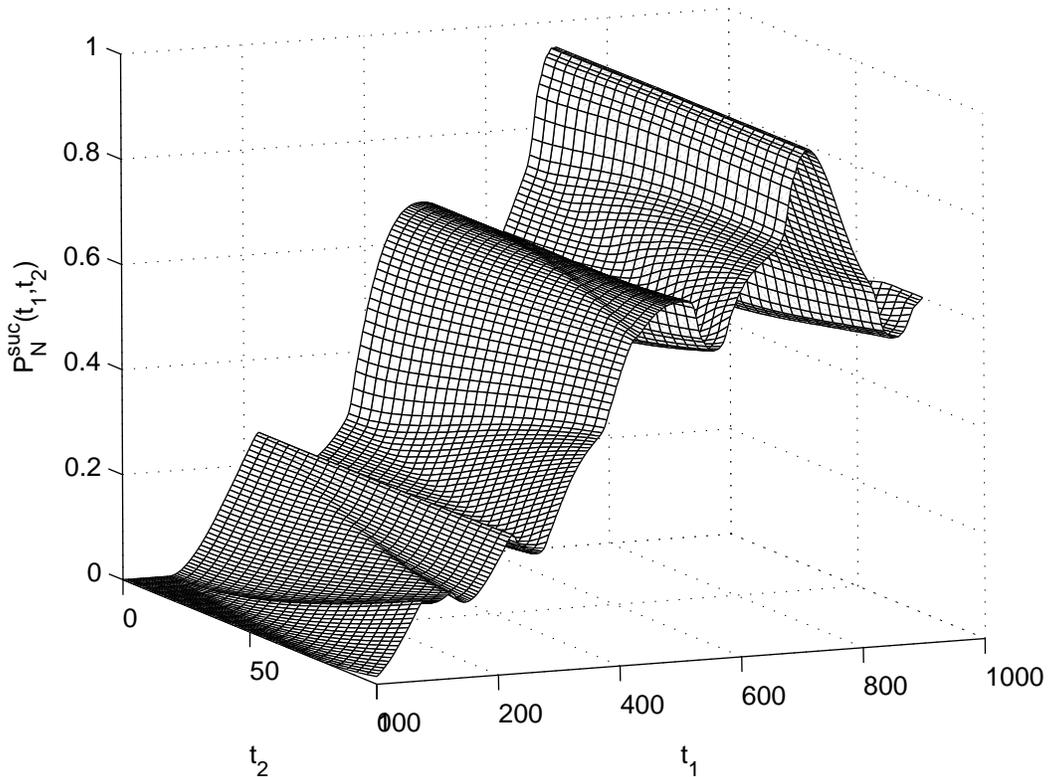}
\caption{\label{fig-5}The dependence of the joint probability of
success, $P_N^{suc}(t_1,t_2)$, on the parameters $t_1$ and $t_2$ is
plotted for $N=150$ and the parameter of $a$ is set to be
\textsl{0.05}. The scales of $t_1$ and $t_2$ are [0 900]
[$\frac{\hbar}{J}$] and [0 100] [$\frac{\hbar}{J}$], respectively. }
\end{figure}

In the following, we will show the dependence of the joint
probability of success $P_N^{suc}(t_1,t_2)$ on the values of $t_1$
and $t_2$, where $t_1$ is the time performing the first measurement,
and $t_2$ is the time interval between the first measurement and the
second one. After straightforward calculation, one can get
\begin{equation}
P_N^{suc}(t_1,t_2)=|{f'}_{N,1}(t_1)|^2+|{f'}_{N,1}(t_1+t_2)-{f'}_{N,N}(t_2){f'}_{N,1}(t_1)|^2,
\label{eq:p2}
\end{equation}
where $P_N^{suc}(t_1,t_2)$ is the joint probability of success. In
Fig.\ref{fig-5}, we plotted the the joint probability of success
$P_N^{suc}(t_1,t_2)$ as a function of $t_1$ and $t_2$ for $N=150$
and $a=0.05$.
In Fig. \ref{fig-5}, one would find that $P_N^{suc}(t_1,t_2)$ is
mainly determined by $t_1$. This is because that the contribution of
the first term in Eq. \ref{eq:p2} to the probability of success is
extremely greater than that of the second term in Eq. \ref{eq:p2}.
Since we want to minimize the time of transferring process and
maximize the $P_N^{suc}(t_1,t_2)$, in turn, the probability of
success for the first measurement will be very large. Therefore, the
significance of the second measurement will be limited. In
Ref.\cite{parallel}, it was proved that the joint probability of
failure would decrease with the increase of the number of
measurements but be impossible to achieve $0$. So we can conclude
that the significance of extra measurements will be even limited,
and by considering the undesired quantum noises that would be
introduced by applying these extra measurements, we only allow two
measurements in the improved protocol.

\begin{figure}
\includegraphics{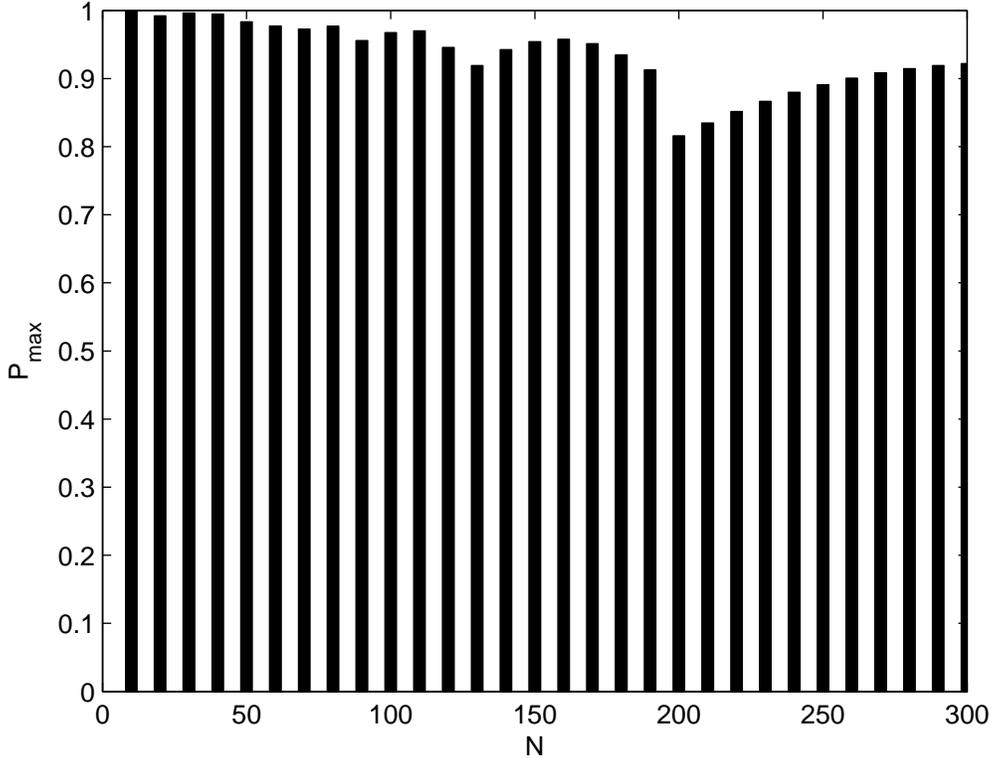}
\caption{\label{fig-P-N}The dependence of $P_{max}$ on the length of
the channel $N$ is plotted. The scales of $t_1$ and $t_2$ are [0
900] [$\frac{\hbar}{J}$] and [0 100] [$\frac{\hbar}{J}$],
respectively. The parameter $a$ is fixed on \textsl{0.05}.}
\end{figure}

If we choose the values of $t_1$ and $t_2$ that can maximize
$P_N^{suc}(t_1,t_2)$, the maximum of the joint probability of
success, denoted by $P_{max}$, will be a function of $N$ and $a$. To
study the dependence of $P_{max}$ on $N$, we fix the parameter $a$.
As an example, we set $a=0.05$. In Fig. \ref{fig-P-N}, we plot the
relationship between $P_{max}$ and $N$ for $a=0.05$. From Fig.
\ref{fig-P-N}, we can see that the maximum of joint probability of
success, $P_{max}$ dose not monotonously decrease with $N$ but
oscillationally decrease with $N$, and that it is not sensitive to
the variation of $N$. Even though $P_{max}$ can not achieve
\textsl{90\%} for some particular $N$, such as $N=200$, one can make
$P_{max}$ achieve \textsl{90\%} by adjusting the parameter $a$ (in
Tab. \ref{tab}).

The requirement of no more than two measurements can also make the
probability of success for perfect states transfer achieve about
\textsl{90\%} and the transferring time less than \textsl{0.1}ns.
For example, when $N=150$, $a=0.05$, $t_1=709$ and $t_2=100$, the
$P_N^{suc}(t_1,t_2)$ can achieve more than \textsl{95\%}, while the
original protocol will take \textsl{1.3}ns to achieve \textsl{90\%}.
What we are interested in is to find some special values of $a$,
where $P_N^{suc}(t_1,t_2)$ can be maximized for reasonable time
scale in experiment. The numerical results of special cases are
listed in Tab.\ref{tab}.
\begin{table}
\caption{\label{tab}The special cases which we are interested in,
for $N=150,200,250,300$, are listed in this table. The time scales
of $t_1$ and $t_2$ we consider are [0 900] [$\frac{\hbar}{J}$] and
[0 100] [$\frac{\hbar}{J}$], respectively. The dimension of time is
[$\frac{\hbar}{J}$], where $J$ is the coupling constant. If $J$ were
taken to be $20K{\times}k_B$, the dimension of time would be
$10^{-4}$ns.}
\begin{ruledtabular}
\begin{tabular}{|c|c|c|c|c|c|c|c|c|c|c|c|c|c|}
\hline
 N                 &\multicolumn{5}{c}{150}\vline
&\multicolumn{3}{c}{200}\vline &\multicolumn{3}{c}{250}\vline &
\multicolumn{2}{c}{300}\vline\\
\hline
$a$&0.05&0.07&0.08&0.14&$0.41\sim0.49$&0.06&0.07&$0.4\sim0.45$&0.05&0.06&$0.36\sim0.44$&0.05&$0.36\sim0.42$\\
\hline
$t_1$&709&411&395&839&81&548&521&106&705&668&131&821&157\\
\hline
$t_2$&100&49&24&59&7&73&100&8&100&44&8&88&9\\
\hline
$P_N^{suc}$&0.95&0.92&0.9&0.87&0.88&0.92&0.88&0.88&0.9&0.91&0.86&0.92&0.86\\
\hline
\end{tabular}
\end{ruledtabular}
\end{table}
The numerical results in Tab.\ref{tab} show that even for large $N$,
the joint probability of success for perfect quantum states transfer
can reach about \textsl{90\%}, which is reasonable in experiment.
Another fact in Tab. \ref{tab} we want to stress is that when the
parameter $a$ is not so small (in the region around \textsl{0.4}),
even though the joint probability of success for perfect transfer is
slightly smaller than the case when a is very small (around
\textsl{0.1}), the total transferring time is significantly reduced.
This fact can make the improved protocol be adapted to different
experimental settings. For example, if the couplings of some system
can not be modulated too much, or the decoherence time of other
system is very short, the experimental groups can select the
specific region they would like to act in. Therefore, the improved
protocol can not only improve the efficiency of quantum-state
transfer, but also adapt to different experimental settings.
Investigating the effects of imperfections is a very important and
practical problem, so we discuss this problem in the next section.

\section{The effects of imperfections}\label{sec4}
Because the improved protocol still use two identical spin chains to
form a dural-rail quantum channel, it not only owns all of the
advantages of resisting quantum noises, such as decoherences, phase
noises and amplitude damping, as discussed in Ref. \cite{parallel},
but is more robust to these quantum noises than the original
protocol. Because the total time $t_{tot}=t_1+t_2$ of the entire
transfer process is less than \textsl{1000} [$\frac{\hbar}{J}$],
which means that the improved protocol only takes less than 0.1ns to
transfer an arbitrary quantum state perfectly with a probability of
about \textsl{90\%} to succeed. Therefore, the improved protocol can
be more robust to these kinds of quantum noises due to the short
time it would take to complete the quantum task.

The thermal effect on the process of transferring quantum states is
another kind of interesting problems in reality. However, the
thermal excitations can be prevented by applying an uniform strong
magnetic field to the chains \cite{parallel}. Therefore, the thermal
effect can be neglected in this case.

Here, we discuss the effects of two kind of specific imperfection on
the probability of success. One kind of imperfection is the
couplings fluctuations caused by the limit of experimental
technology of engineering spin chains. Another kind of imperfection
is caused by imperfect initialization of the spin chains. Due to the
limitation of experimental technology of initializing the spin
system, this kind of imperfection can happen in the process of
initializing the spin chains. Therefore, discussing the effects of
these two kinds of imperfections on both the improved protocol and
the original one has much significance. In the following, we will
show that the improved protocol is more robust to the two kinds of
imperfections than the original one.

\subsection{The effect of imperfect couplings}
In Ref. \cite{fluc2}, Daniel Burgarth and Sougato Bose studied the
effect of couplings fluctuations introduced by the interaction
between the spin chain and the spin bath, to the state transfer in
single spin-chain channel. The results are interesting. Here, we
discuss the effect of imperfect couplings caused by the limitation
of experimental technology, such as the accuracy of distance between
adjacent spins and the fluctuations of intermediate material's
density, to the dual-rail spin-chain channels. To show the
advantages of the improved scheme, we also compare the effects of
imperfect couplings to the two schemes.

Because of the limit of experimental technology of engineering spin
chains, it is impossible to engineer each of the couplings to the
theoretical value, so the couplings fluctuation is very common. As
we discussed above, both of the improved scheme and the original one
require two identical spin chains, so someone may believe that the
slight difference between the two spin chains caused by couplings
fluctuation will be lethal to the scheme of dual-rail quantum
channels. However, this is not the case \cite{fluc}. In Ref.
\cite{fluc}, the authors showed that one can also accomplish the
conclusive transfer with two different spin chains.

In this section, we show the dependence of the probability of
success for perfect transfer on the parameter $a$ with considering
the couplings fluctuations, so that the advantage of the improved
protocol can be displayed on resisting the couplings fluctuations.

When considering the couplings fluctuations, the Hamiltonians of the
chains are given by
\begin{eqnarray}\label{eq:fluc-1}
H_{\mathrm{inh}}^{(1)}=\frac{a}{2}(1+\delta_1^{(1)})(\sigma_1^x\sigma_2^x+\sigma_1^y\sigma_2^y)
                      +\frac{a}{2}(1+\delta_{N-1}^{(1)})(\sigma_{N-1}^x\sigma_N^x+\sigma_{N-1}^y\sigma_N^y)\nonumber\\
                      +\frac{J}{2}\sum^{N-2}_{n=2}{(1+\delta_n^{(1)})(\sigma_n^x\sigma_{n+1}^x+\sigma_n^y\sigma_{n+1}^y)}
                      +\sum_{n=1}^N{\omega\sigma_n^z},
\end{eqnarray}
\begin{eqnarray}\label{eq:fluc-2}
H_{\mathrm{inh}}^{(2)}=\frac{a}{2}(1+\delta_1^{(2)})(\sigma_1^x\sigma_2^x+\sigma_1^y\sigma_2^y)
                      +\frac{a}{2}(1+\delta_{N-1}^{(2)})(\sigma_{N-1}^x\sigma_N^x+\sigma_{N-1}^y\sigma_N^y)\nonumber\\
                      +\frac{J}{2}\sum^{N-2}_{n=2}{(1+\delta_n^{(2)})(\sigma_n^x\sigma_{n+1}^x+\sigma_n^y\sigma_{n+1}^y)}
                      +\sum_{n=1}^N{\omega\sigma_n^z},
\end{eqnarray}
where $\delta^{(i)}_n$ are uniformly distributed uncorrelated random
numbers in the interval $[-\Delta,\Delta]$. It is reasonable to
require the experimental precision $\Delta$ less than $0.01$. In the
following discussion, we just set $\Delta=0.01$.

To illustrate the advantage of the improved protocol on resisting
this kind of quantum noise, only considering the probability of
success after performing the first measurement is enough. Because
the two spin chains are different, each of the spin chains will
evolve respectively after encoded the quantum state to the dual-rail
spin-chain channel. Generally speaking,
${f'}_{N,1}^{(1)}(t)\neq{f'}_{N,1}^{(2)}(t),$ where
${f'}_{N,1}^{(1)}(t)=\langle\bm{N}|e^{-iH^{(1)}_{\mathrm{inh}}t}|\mathbf{1}\rangle$
and
${f'}_{N,1}^{(2)}(t)=\langle\bm{N}|e^{-iH^{(2)}_{\mathrm{inh}}t}|\mathbf{1}\rangle.$
However, $|{f'}_{N,1}^{(1)}(t)|$ intersects with
$|{f'}_{N,1}^{(2)}(t)|$ many times \cite{fluc}, so to make the
receiver's measurements unbiased with respect to the initial state
(necessary requirement for conclusive transfer), the receiver has to
perform CNOT operation and measurement to the \textsl{N}th site of
spins at time $\tau$ \cite{fluc}, where
\begin{equation}\label{eq:case}
|{f'}_{N,1}^{(1)}(\tau)|=|{f'}_{N,1}^{(2)}(\tau)|.
\end{equation}
Because the phase of ${f'}_{N,1}^{(1)}(\tau)$ is different with that
of ${f'}_{N,1}^{(2)}(\tau)$, when the receiver obtain ``1" after the
measurement, he has to apply a phase operation to correct the phase
error.

\begin{figure}
\includegraphics{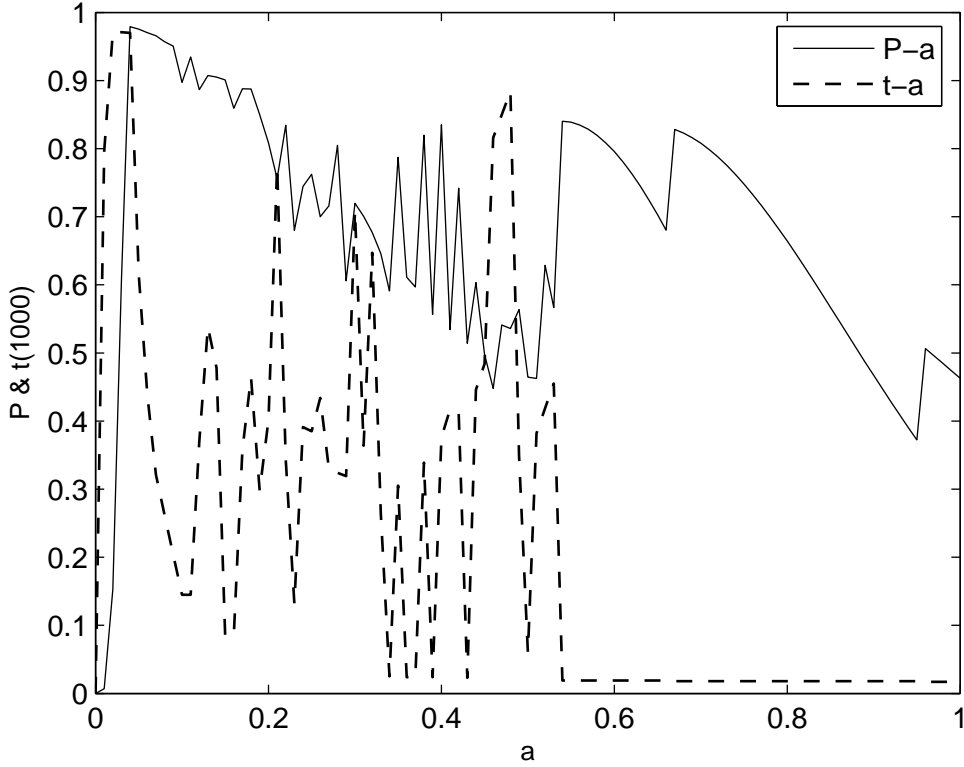} \caption{\label{fig-11} The dependence of the maximal probability of
success (solid line) and of the times at which the maximal
probability of success (dashed line) is achieved on the parameter
$a$ for $N=30$. The time scale where we search the points of
intersection is set to be less than $1000[\frac{\hbar}{J}].$ The
samples of fluctuations, \{$\delta^{(1)}_n$\} and
\{$\delta^{(2)}_n$\}, are chosen to be \{0.0091, 0.0031, 0.0048,
-0.0031, 0.0077, -0.0031, -0.0088, 0.0044, 0.0092, -0.0069, -0.0017,
-0.0081, -0.0010, 0.0074, -0.0022, -0.0049, -0.0029, 0.0049, 0.0030,
0.0088, 0.0067, -0.0006, 0.0026, -0.0088, 0.0008, -0.0009, 0.0073,
0.0071, -0.0006\} and \{0.0057, 0.0031, -0.0100, -0.0074, -0.0001,
-0.0092, -0.0055, -0.0034, 0.0080, -0.0037, -0.0050, -0.0013,
0.0068, -0.0063, 0.0002, -0.0010, -0.0035, -0.0024, 0.0077, 0.0052,
0.0077, -0.0009, 0.0060, -0.0073, -0.0087, -0.0025, -0.0025,
-0.0003, 0.0094\}, respectively, which are generated by computer.}
\end{figure}

We first search specific cases that satisfy Eq. (\ref{eq:case}), for
$0<a<1$ and $0<t<1000$, then find out the maximal probability of
success for specific value of $a$. In Fig. \ref{fig-11}, we show the
relationship between the maximal probability of success and the
parameter $a$, and we also plot dependence of the times at which the
maximal probability of success is achieved on the parameter $a$.

From Fig. \ref{fig-11}, we can see that one can also achieve high
probability of success by modulating the parameter $a$, even
considered the couplings fluctuations, and it is possible to take
little time to achieve high probability of success for some specific
values of $a$, such as $a=0.11, 0.54$. To accomplish the conclusive
transfer, one need to determine the modulus of ${f'}_{N,1}^{(1)}(t)$
and ${f'}_{N,1}^{(2)}(t)$ by the method of tomography \cite{fluc},
rather than precisely measuring every coupling. The precise
requirement to the times at which the receiver should measure can be
relaxed by measuring at times where not only the probability
amplitudes are similar, but also their slope \cite{fluc}.

\begin{figure}
\includegraphics{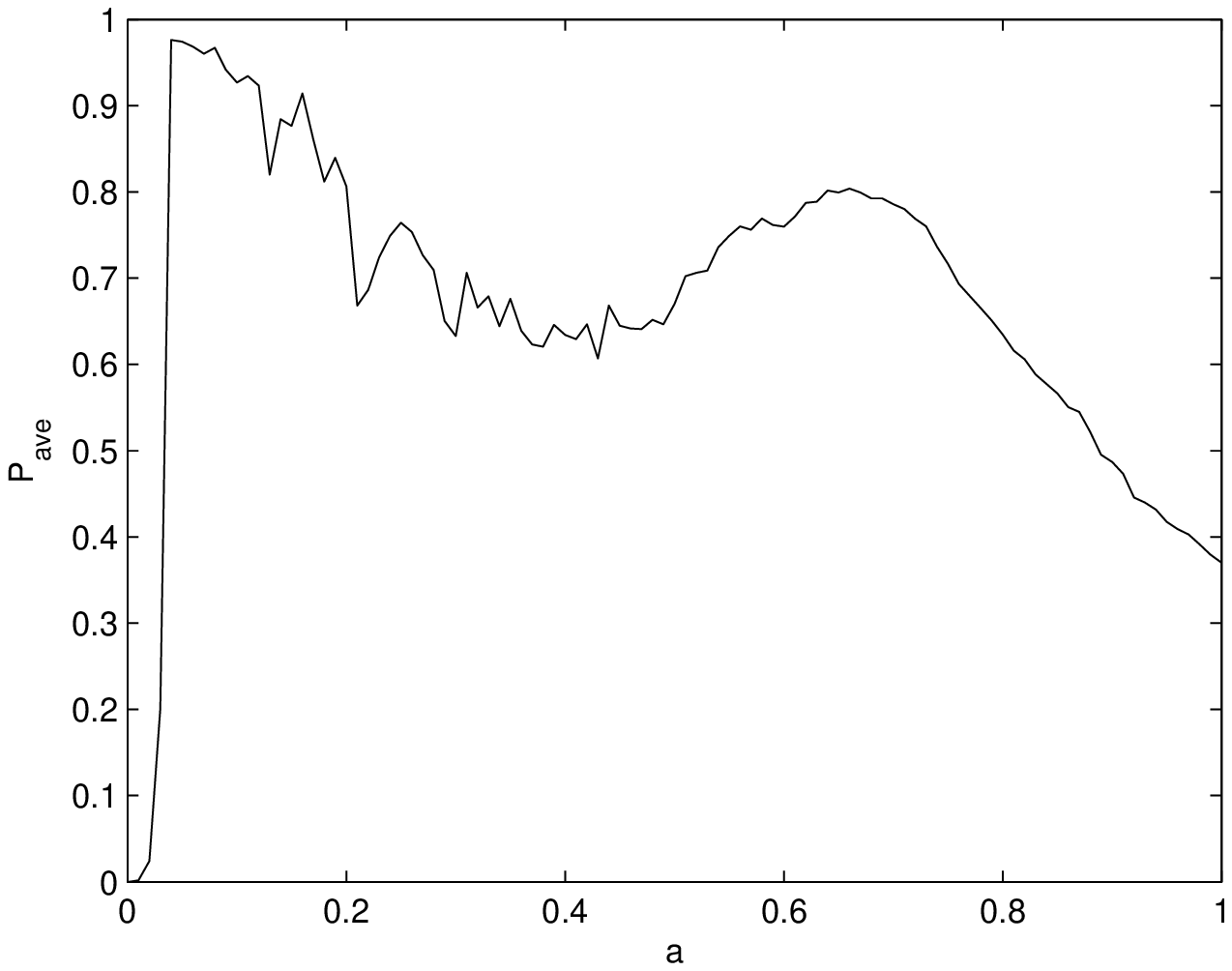}
\caption{\label{fig-Pave20}We study $20$ random samples for $N=30$
and calculate the average probability of success, $P_{ave}$. The
dependence of $P_{ave}$ on the parameter $a$ is plotted. The times
are chosen to maximize the probability of success for each sample.
All of the random numbers are generated by computer in the interval
$[-\Delta,\Delta]$, where $\Delta=0.01$.}
\end{figure}

Even though the advantages of the improved scheme can be displayed
in the specific example (Fig. \ref{fig-11}), we also examined $20$
sets of random samples of fluctuations and calculated the average
probability of success. The dependence of the average probability of
success, $P_{ave}$, on the parameter $a$ is plotted in Fig.
\ref{fig-Pave20}. The results showed us that by modulating the
parameter $a$, one can significantly improve the probability of
success.

\subsection{The effect of imperfect initialization}
The imperfections caused by imperfect initialization of the system
are common in reality, because memory effects and defects usually
exist in quantum operation systems. In the following sections, we
discuss two specific cases of this kind of imperfection, random
single excitation and collective excitation.
\subsubsection{The imperfection of probabilistic single excitation}
By random single excitation we mean that only one excitation happens
in each spin chain due to the imperfect initialization caused by the
defect of experimental settings. We assume the two excitations
happen on the same site of the two spin chains, respectively. The
reason that we make such an assumption is that the experimental
environments with which the two spin chains confront are the same.
When we consider this kind of noise, the initialized state of the
system would be:
\begin{eqnarray}
|\phi\rangle&=&(\sqrt{1-x}|\mathbf{0}\rangle^{(1)}+\sqrt{x}|\bm{m}\rangle^{(1)})\otimes(\sqrt{1-x}|\mathbf{0}\rangle^{(2)}+\sqrt{x}|\bm{m}\rangle^{(2)}) \nonumber\\
            &=&(1-x)|\mathbf{0}\rangle^{(1)}|\mathbf{0}\rangle^{(2)}+\sqrt{x(1-x)}(|\mathbf{0}\rangle^{(1)}|\bm{m}\rangle^{(2)}+|\bm{m}\rangle^{(1)}|\mathbf{0}\rangle^{(2)})+x|\bm{m}\rangle^{(1)}|\bm{m}\rangle^{(2)}
            \label{eq:random}
\end{eqnarray}
where $x$ is the probability of happening the random excitation, and
$m$ is the site where excitation happens. The second term of Eq.
(\ref{eq:random}) represents the situation that the two spin chains
are in different states. After encoding the quantum state to the
channel, the evolution of spin chain (1) in this term is different
(in either amplitude or phase) from that of spin chain (2). Because
the main contribution to the probability of success is from the
first term of Eq. (\ref{eq:random}), one will either not choose the
special times when the amplitudes of spin chain (1) and (2) are
intersecting with each other, nor perform phase correction.
Therefore, this term will have no contribution to the probability of
success. After encoding the quantum state to be transferred to the
first sites of the dual-rail quantum channels and applying
corresponding unitary transformation described in the protocol, the
state of the system will be
$$ |\psi(0)\rangle=(1-x)|s(1)\rangle+x(\alpha|\bm{m}\rangle^{(1)}|\bm{1m}\rangle^{(2)}
+\beta|\bm{1m}\rangle^{(1)}|\bm{m}\rangle^{(2)}),$$ where
$|\bm{1m}\rangle$ represents the state of the system with the qubits
of site 1 and site $m$ in excited states and all other qubits in
ground states. The number of excitations in any single spin chain
will conserve in the time evolution, since
$[\sigma^z_{tot},H_{\mathrm{inh}}]=0.$ So after time \textsl{t}, the
state of the system would be
\begin{equation}\label{eq:noise-t}
|\psi(t)\rangle=(1-x)\sum^N_{n=1}{{f'}_{n,1}(t)|s(n)\rangle}
+x\sum^N_{n=1}{\sum_{pq}{{f'}_{n,m}(t){f'}_{1m}^{pq}(t)(\alpha|\bm{n}\rangle^{(1)}|\bm{pq}\rangle^{(2)}
+\beta|\bm{pq}\rangle^{(1)}|\bm{n}\rangle^{(2)})}},
\end{equation}
where
${f'}_{n,m}(t)=\langle\bm{n}|e^{-iH_{\mathrm{inh}}t}|\bm{m}\rangle,$
and
${f'}^{pq}_{1m}(t)=\langle\bm{pq}|e^{-iH_{\mathrm{inh}}t}|\bm{1m}\rangle.$
The second term of Eq. (\ref{eq:noise-t}) can be rewrite as:
\begin{eqnarray}\label{eq:noise-t2}
&&\sum^{N-1}_{n=1}{\sum_{p<q \atop q<N}{{f'}_{n,m}(t){f'}^{pq}_{1m}(\alpha|\bm{n}\rangle^{(1)}|\bm{pq}\rangle^{(2)}+\beta|\bm{pq}\rangle^{(1)}|\bm{n}\rangle^{(2)})}}\nonumber\\
&&+\sum_{p<q \atop q=N}{{f'}_{N,m}(t){f'}_{1m}^{pN}(t)(\alpha|\bm{N}\rangle^{(1)}|\bm{pN}\rangle^{(2)}+\beta|\bm{pN}\rangle^{(1)}|\bm{N}\rangle^{(2)})}\nonumber\\
&&+\sum^{N-1}_{n=1}{\sum_{p<q \atop q=N}{{f'}_{n,m}(t){f'}_{1m}^{pN}(t)(\alpha|\bm{n}\rangle^{(1)}|\bm{pN}\rangle^{(2)}+\beta|\bm{pN}\rangle^{(1)}|\bm{n}\rangle^{(2)})}}\nonumber\\
&&+\sum_{p<q \atop
q<N}{{f'}_{N,m}(t){f'}_{1m}^{pq}(t)(\alpha|\bm{N}\rangle^{(1)}|\bm{pq}\rangle^{(2)}+\beta|\bm{pq}\rangle^{(1)}|\bm{N}\rangle^{(2)})}
\end{eqnarray}
After performing the C-NOT gate on the \textsl{N}th site of the
dual-rail quantum channel, the measurement of the \textsl{N}th qubit
of the spin chain (2) would be ``0" for the first and second terms
of Eq. (\ref{eq:noise-t2}) and ``1" for the third and forth terms of
Eq. (\ref{eq:noise-t2}). The result of ``0" represents the failure
of transfer, which is one of the advantages of dual-rail quantum
channels. However, the result of ``1" represents both perfect
transfer and imperfect transfer, which means that this kind of
probabilistic single excitation destroys the greatest
advantage---conclusive transfer. For the third term of Eq.
(\ref{eq:noise-t2}), only when $n=p$, the result of ``1" means
perfect transfer. While all other terms with measurement of ``1" but
$n{\neq}p$, the \textsl{N}th qubit of the spin chain (1) would be in
the state of
$\rho=|\alpha|^2|0\rangle\langle0|+|\beta|^2|1\rangle\langle1|,$
which is a mixed state. Therefore, the probability of success of
perfect transfer for Eq. (\ref{eq:noise-t}) will be
\begin{equation}\label{eq:noise-pro}
P^{suc}(t)=(1-x)^2P^{suc}_0(t)+x^2P^{suc}_1,
\end{equation}
where $P^{suc}_0=|{f'}_{N,1}(t)|^2$ and
$P^{suc}_1=\sum^{N-1}_{n=1}{|{f'}_{n,m}(t)|^2|{f'}_{1m}^{nN}(t)|^2}.$

We just consider the situation with only one measurement permitted,
for two reasons, one is that further measurements would introduce
more probability of imperfect transfer, and the other one is that
the probability of success for perfect transfer is mainly determined
by the probability of the result of the first measurement being
``1"---$P^{suc}_0(t)$, as discussed above. By considering the
situation with only one measurement, the ability to resist this kind
of imperfection can be illustrated for the improved protocol. In the
following, we study this problem for $N=30$, the results of which
would be applicable to even larger length.

We are interested with the probability of success for perfect
transfer $P^{suc}(t)$. This quantity is related with parameters of
$a$, $t$, $m$ and $x$. Generally speaking, one can make the
probability of happening this kind of single excitation extremely
small, but not eliminate it, it would be reasonable to set $x\le0.1$
when one study the relationship between the two interesting
quantities and the other three parameters. Since the parameter $x$
is small, the probability of success for perfect transfer is mainly
determined by the $P^{suc}_0$. One can set the parameters of $a$ and
$t$ to the values that would maximize $P^{suc}_0$. It would be
interesting to study the relationship between the probability of
$P^{suc}_1$ and the parameter $m$, which is the site where the
random noise happens.
\begin{figure}
\includegraphics{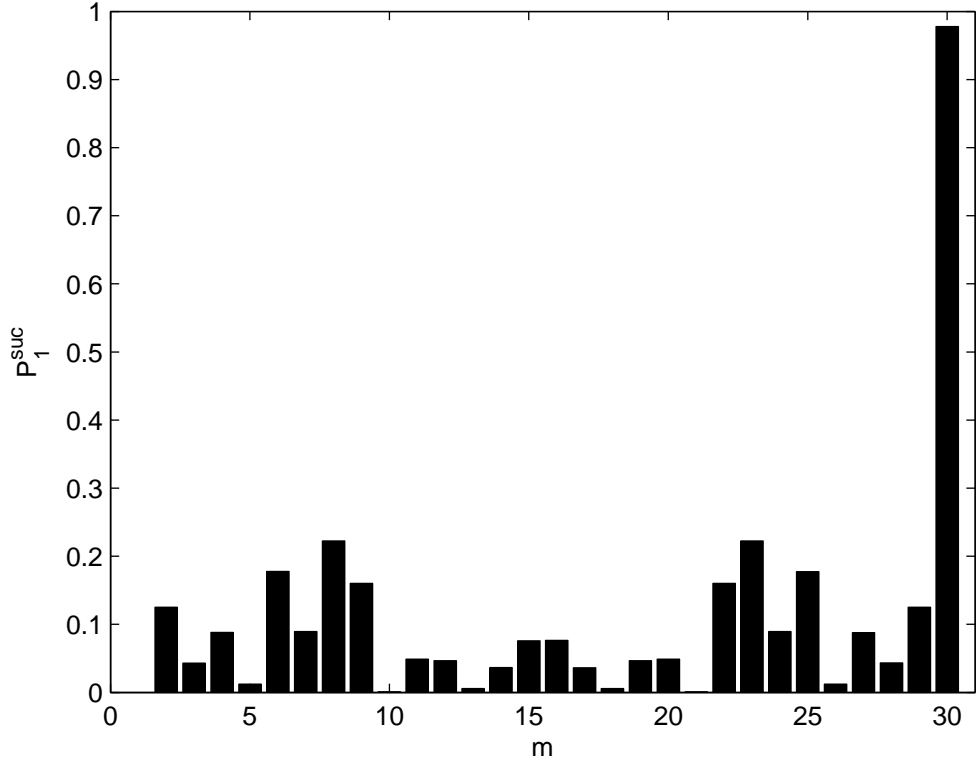} \caption{\label{fig-6}The
dependence of $P^{suc}_1$ on the parameter $m$ is displayed in the
figure, for $N=30$,$a=0.06$ and $t=488[\frac{\hbar}{J}]$.}
\end{figure}
From Fig. \ref{fig-6}, we can see that the values of $P^{suc}_1$ are
significantly depended on the site where the qubit is in excited
state caused by the imperfect initialization. When the noises happen
in some particular sites, such as \textsl{5, 10, 13, 21}, the values
of $P^{suc}_1$ are approximately equal to zero. So if one can take
some measures to prevent the single excitation to happen on these
particular sites, he or she can certainly improve the probability of
success for perfect transfer. Another feature of Fig. \ref{fig-6} is
that the value of $P^{suc}_1$ is extremely large when $m=N$. This is
because when $m=N$, part of the initial state after encoding the
quantum state to be transferred to the first site of the quantum
channel, is $|\bm{1m}\rangle=|\bm{1N}\rangle$, which is mirror
symmetric about its center. This phenomena directly testify that
mirror symmetry is very important to improve the transferring
efficiency.

Generally speaking, we do not know which site of the dual-rail
quantum channel will be in excited state. Therefore, it is common to
assume that the probability of being in excited state for each site
is equal since the environment confronted by the two spin chains is
uniform. It would be practical and interesting to study the
dependence of the average probability of success for perfect
transfer on the parameter $a$, when the transferring time was chosen
to maximize the average probability of success for perfect transfer.
\begin{figure}
\includegraphics{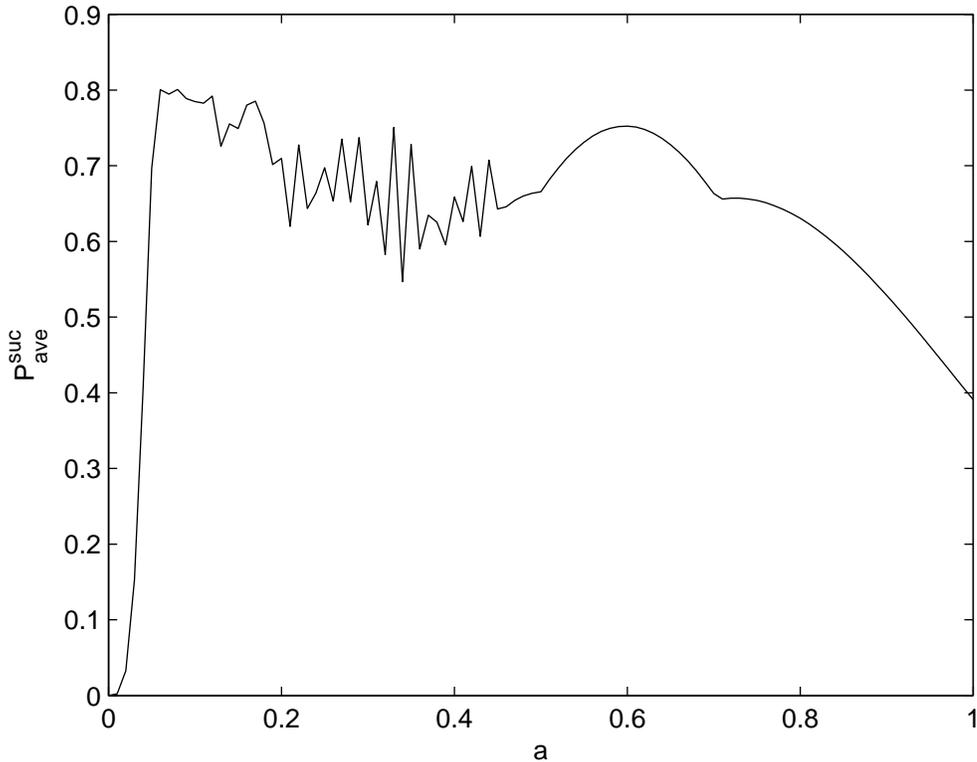}
\caption{\label{fig-7}In this figure, we plot the dependence of the
average probability of success for perfect transfer $P_{ave}^{suc}$
on the parameter $a$ for $N=30$. The transferring time for each $a$
was set to be the values that would maximize $P^{suc}_{ave}$ and the
value of $x$ was set to be \textsl{0.1}.}
\end{figure}
In the Fig. \ref{fig-7}, we can see that the average probability of
success for perfect transfer $P_{ave}^{suc}$ is significantly
affected by the changes of the parameter $a$, therefore, one can
improve the average probability of success for perfect transfer by
adjusting the parameter $a$. Numerical calculation shows that when
$a=0.06$, the $P^{suc}_{ave}$ can be maximized and the maximum of
the $P^{suc}_{ave}$ is \textsl{0.80}, while the maximum of
$P^{suc}_{ave}$ is only \textsl{0.39} when $a=1$, which represent
the situation of original protocol. Apparently, The probability of
success for perfect transfer $P^{suc}_{ave}$ is a quadratic function
of $x$. To completely understand the advantages of the improved
protocol, we need to examine the exact relationship between
$P^{suc}_{ave}$ and the parameter $x$ for the original protocol and
the improved protocol, respectively.
\begin{figure}
\includegraphics{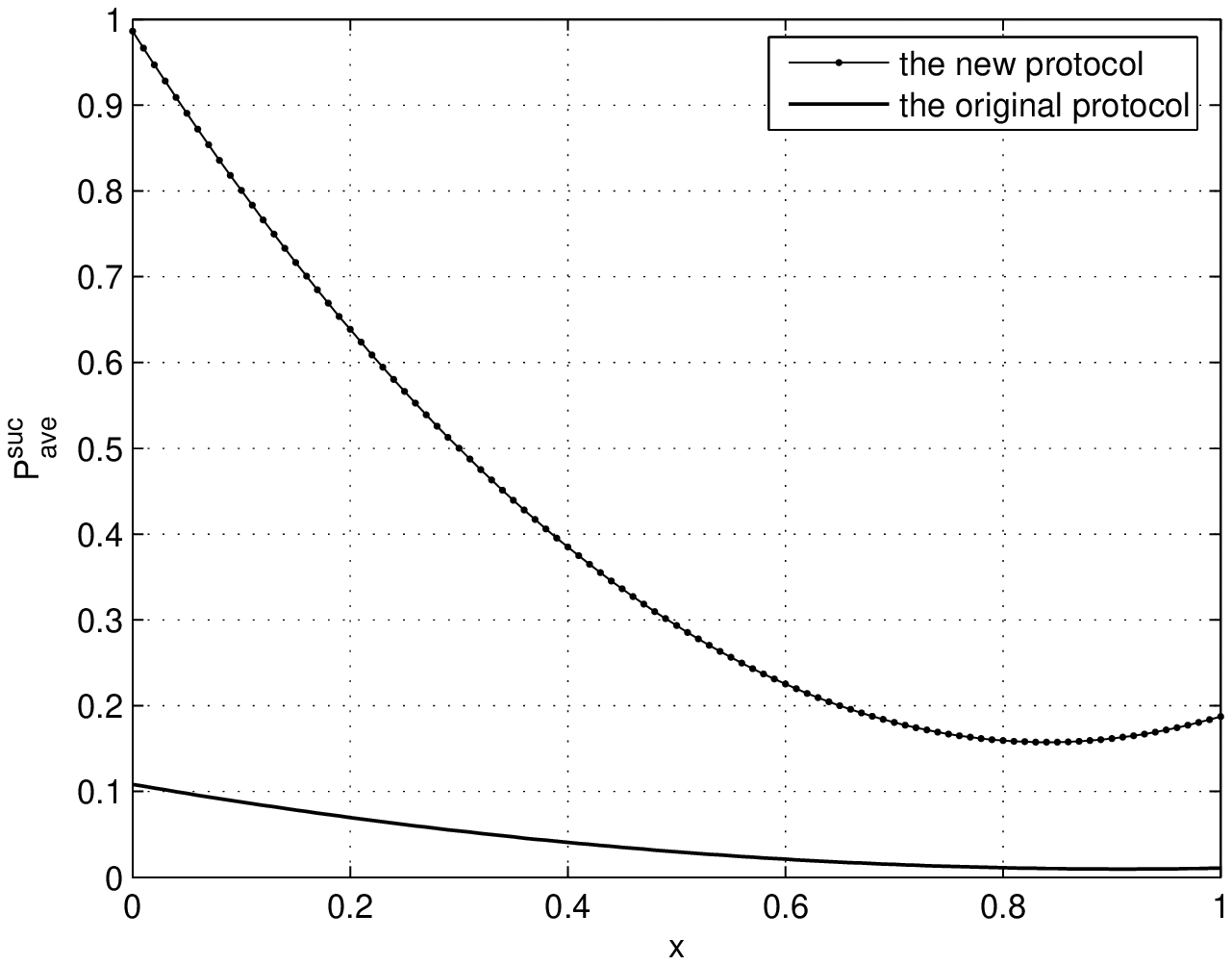}
\caption{\label{fig-8}} In this figure, we plot the dependence of
the average probability of success for perfect transfer
$P_{ave}^{suc}$ on the parameter $x$, for the improved protocol (the
solid line with dots) and the original protocol (the solid line).
For the improved protocol, we set the parameter $a=0.06$ and
$t=488$, which can maximize $P_{ave}^{suc}$ for $N=30$.
\end{figure}
In Fig. \ref{fig-8}, we can see the advantage of the improved
protocol on resisting the probabilistic single excitation.
Therefore, the improved protocol will be more robust than the
original one to this kind of imperfection, which is introduced by
imperfect initialization.

\subsubsection{The imperfection of probabilistic collective excitation}
Collective excitation is another common type of imperfection that
can happen in spin-chain channels, because this kind of imperfecion
can also be caused by imperfect initialization resulting from the
limitation of experimental settings. Therefore, addressing the
collective excitation in this improved scheme has much practical
significance.

\begin{figure}
\includegraphics{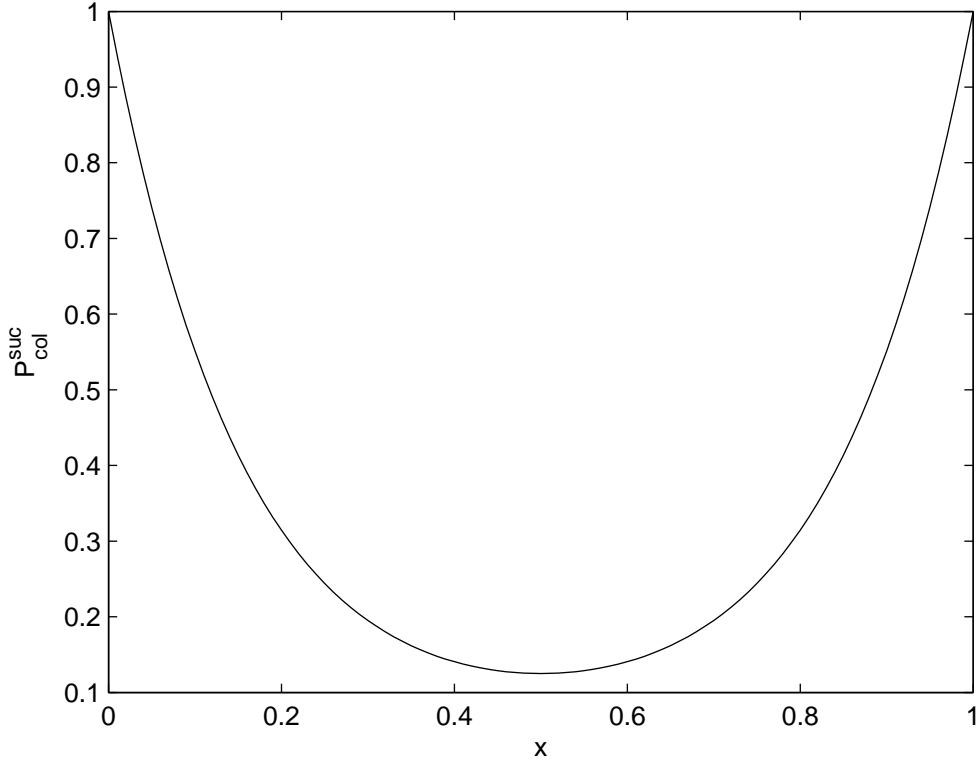}
\caption{\label{fig-9} The dependence of $P_{col}^{suc}$ on the
parameter $x$ is plotted for $N=4$. We set $a=0.06$ and $t=18$,
which will maximize the $P_{col}^{suc}$.}
\end{figure}

We just assume the first qubits of the dual-rail spin-chain channel
was cooled down to their ground state, since the procedure of
encoding the prepared state to the channel will completely eliminate
the effect of the imperfecition on the first qubits. After
considering the collective excitation caused by imperfect
initialization, the system, in fact, will be initialized to the
state
\begin{equation}\label{eq:col-noise}
|\phi\rangle=|0_{1}\rangle^{(1)}\otimes\prod_{i=2}^{N}{(\sqrt{1-x}|0_i\rangle^{(1)}+\sqrt{x}|1_i\rangle^{(1)})}
\otimes|0_1\rangle^{(2)}\otimes\prod_{j=2}^N{(\sqrt{1-x}|0_j\rangle^{(2)}+\sqrt{x}|1_j\rangle^{(2)})},
\end{equation}
where $x$ represents the probability of each qubit being in excited
state, and $|0_k\rangle^{(c)}$ and $|1_k\rangle^{(c)}$ represents
that the \textsl{k}th qubit of spin chain (c) $(c=1,2)$ is in its
ground state and excited state, respectively. When we expand the Eq.
(\ref{eq:col-noise}), we would find that it is not necessary to
analyze all of the terms, because the terms that represent that the
spin chain (1) and the spin chain (2) are in different states have
no contribution to the probability of success (for the same reason
as discussed in the subsection of probabilistic single excitation).
Therefore, we just consider the terms that will contribute to the
probability of success. After encoding the quantum state
$|\psi\rangle$ to the first site of spin chain (1) and applying the
similar C-NOT operation controlled by the first site of spin chain
(1) being ``0" to the first sites of the dural-rail channels, the
contributing part of the system's state will become
\begin{eqnarray}\label{eq:col-expand}
|\psi(0)\rangle&=&(1-x)^{N-1}(\alpha|\mathbf{0}\rangle^{(1)}|\mathbf{1}\rangle^{(2)}+\beta|\mathbf{1}\rangle^{(1)}|\mathbf{0}\rangle^{(2)})\nonumber\\
               &&+x(1-x)^{N-2}\sum_{m=2}^N{(\alpha|\bm{m}\rangle^{(1)}|\bm{1m}\rangle^{(2)}+\beta|\bm{1m}\rangle^{{1}}|\bm{m}\rangle^{(2)})}\nonumber\\
               &&+x^2(1-x)^{N-3}\sum_{n=2,p=3 \atop n<p}^{n=(N-1), \atop
               p=N}{(\alpha|\bm{np}\rangle^{(1)}|\bm{1np}\rangle^{(2)}+\beta|\bm{1np}\rangle^{(1)}|\bm{np}\rangle^{(2)})}\nonumber\\
               &&+\cdots \nonumber\\
               &&+x^{N-1}(\alpha|\bm{23\cdots N}\rangle^{(1)}|\bm{123\cdots N}\rangle^{(2)}+\beta|\bm{123\cdots
               N}\rangle^{(1)}|\bm{23\cdots
               N}\rangle^{(2)}),
\end{eqnarray}
where $|\bm{ij\cdots k}\rangle (i<j<\cdots<k)$ represents the state
that \textsl{i}th qubit, \textsl{j}th qubit,$\cdots$, and
\textsl{k}th qubit are in their excited states, while others are in
their ground states. Since the number of excitations are conserved
under free evolution, after time $t$, the state will evolve to
\begin{equation}
\begin{array}{ll}
|\psi(t)\rangle=&(1-x)^{N-1}\sum_{i=1}^n{f'_{i,1}(t)(\alpha|\mathbf{0}\rangle^{(1)}|\mathbf{i}\rangle^{(2)}+\beta|\mathbf{i}\rangle^{(1)}|\mathbf{0}\rangle^{(2)})}\\
               &+x(1-x)^{N-2}\sum_{m=2}^N{\sum_{j=1}^N{\sum_{r<s}^{}{f'_{j,m}(t){f'}_{1m}^{rs}(t)(\alpha|\mathbf{j}\rangle^{(1)}|\mathbf{rs}\rangle^{(2)}+\beta|\mathbf{rs}\rangle^{(1)}|\mathbf{j}\rangle^{(2)})}}}\\
               &+\cdots \qquad \cdots \\
               &+x^{N-1}\sum_{u_1<\cdots <u_{N-1}}{{f'}_{23\cdots N}^{u_1\cdots u_{N-1}}(t)}\times\\
               &\qquad\qquad(\alpha|\bm{u_1\cdots u_{N-1}}\rangle^{(1)}|\bm{123\cdots N}\rangle^{(2)}
               +\beta|\bm{123\cdots N}\rangle^{(1)}|\bm{u_1\cdots u_{N-1}}\rangle^{(2)}),
\end{array}\label{eq:col-evolve}
\end{equation}
we can analyze each term of Eq. (\ref{eq:col-evolve}) as we analyze
Eq. (\ref{eq:noise-t}). After complicated analysis, we can get the
probability of success as
\begin{equation}
\begin{array}{ll}
P^{suc}_{col}(t)=&(1-x)^{2(N-1)}|{f'}_{N,1}(t)|^2+x^2(1-x)^{2(N-2)}\sum_{m=2}^{N}{\sum_{j=1}^{N-1}{|{f'}_{j,m}(t)|^2|{f'}_{1m}^{jN}(t)|^2}}\\
&+x^4(1-x)^{2(N-3)}\sum_{n=2,p=3 \atop n<p}^{p=N, \atop
n=(N-1)}{\sum_{r<s}^{s=N-1, \atop
r=N-2}{|{f'}_{np}^{rs}(t)|^2|{f'}_{1np}^{rsN}(t)|^2}}+\cdots \\
&+x^{2(N-1)}|{f'}_{23\cdots N}^{12\cdots (N-1)}(t)|^2.
\end{array}\label{eq:col-suc}
\end{equation}
From Eq. (\ref{eq:col-suc}), we can see that the probability of
success for perfect transfer $P^{suc}_{col}$ is the $2(N-1)$ order
function of the parameter $x$. The function's shape is determined by
the coefficient of each term. Since calculating the probability of
success $P^{suc}_{col}$ is too difficult for long channels, we just
numerically calculated the result of $P^{suc}_{col}$ for $N=4$. The
exact results is plotted in Fig. \ref{fig-9}. There are two
interesting features in Fig. \ref{fig-9}. The first one is the
symmetry of the curve, which results from the Hamiltonian, because
in the subspace with $\mu$ excitations the Hamiltonian
$H_{\mathrm{inh}}$ has the same matrix as in the subspace with
$N-\mu$ excitations. For example, ${f'}_{n,m}={f'}^{12\cdots
0_n\cdots N}_{12\cdots 0_m\cdots N}$. The other one is that
$P^{suc}_{col}$ decreases raptly with the increase of parameter $x$,
when $x<0.5$. This implies that the effect of collective excitation
is much severe, and the effect will exponentially increase with the
length of the channel $N$.

\begin{figure}
\includegraphics{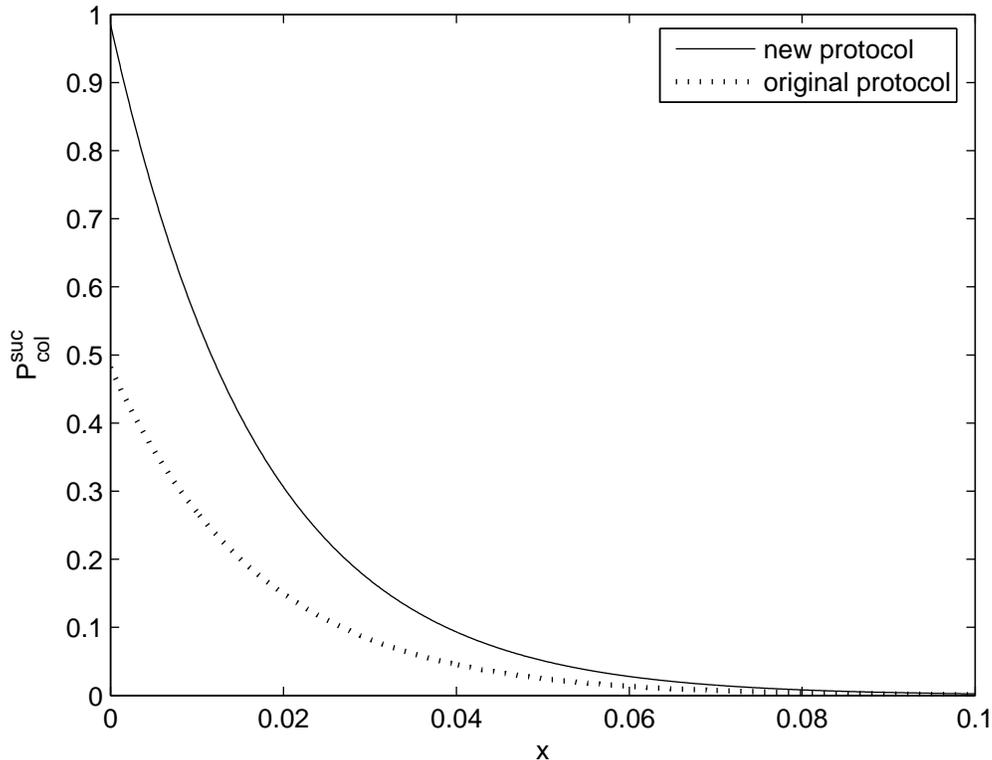}
\caption{\label{fig-10} The effect of collective excitation to the
probability of success $P_{col}^{suc}$ is plotted in this figure for
the improved protocol (solid line) and the original protocol (dotted
line). The length of the channel is chosen to be $N=30$, and the
parameter $a$ is set to be $0.06$ for the improved protocol.}
\end{figure}

In most cases, the probability of happening collective excitation
$x$ is very small, or the spin-chain channels can not complete the
quantum communication effectively. Therefore, it is reasonable to
assume $x<0.1$. As a result, we can ignore all the terms in Eq.
(\ref{eq:col-suc}) except the first one and the second one, because
the terms with higher orders of the parameter $x$ are extremely
small compared with the first two terms. In the case of $x<0.1$, the
probability of success will be
\begin{equation}\label{eq:col-part}
P^{suc}_{col}(t)=(1-x)^{2(N-1)}|{f'}_{N,1}(t)|^2+x^2(1-x)^{2(N-2)}\sum_{m=2}^{N}{\sum_{j=1}^{N-1}{|{f'}_{j,m}(t)|^2|{f'}_{1m}^{jN}(t)|^2}}
\end{equation}
The probability of success $P_{col}^{suc}$ in Eq.
(\ref{eq:col-part}) can be calculated even for long channels. In
Fig. \ref{fig-10}, we numerically calculate the dependence of
$P_{col}^{suc}$ on the parameter $x$ for $N=30$ for both protocols.
From Fig. \ref{fig-10}, we can see the improved protocol is more
robust to collective excitation than the original one.

\section{Conclusion and discussion}\label{sec5}
In this paper, we successfully find a way to avoid the quantum noise
introduced by unnecessary operations for the scheme of
\cite{parallel}. It is to substitute the uniform spin chains used in
the original scheme, with the inhomogenous spin chains studied in
Ref. \cite{endsc}, so that the number of operations and measurements
needed to achieve reasonable probability of success can decrease
significantly. Furthermore, the improved scheme  not only maintains
its naturality and simplicity, but also greatly improved the
efficiency of quantum-state transfer. We carefully studied the
dependence of the probability of success for perfect transfer on the
parameter $a$ and on the length of channel $N$. Results show that
changing the value of the parameter $a$ can not only improve the
probability of success, but also shorten the transferring time, and
that the probability of success dose not monotonously decrease with
$N$ and is not sensitive to the change of $N$. Further, we give out
the special cases that can greatly improve the probability
(achieving about \textsl{90\%}) of success for
\textsl{N}=\textsl{150, 200, 250} and \textsl{300}, which are
considerable large. Finally, we also studied the effects of
couplings fluctuations and imperfect initialization. The results
show that the improved scheme is more robust to these two kinds of
imperfections than the original one. Even the couplings fluctuations
happens, the research results show that the conclusive transfer is
still possible and that the effect of this kind of imperfection is
not lethal to the scheme of dual-rail spin-chain channel. However,
the imperfections of probabilistic single excitation and of
probabilistic collective excitations, caused by imperfect
initialization will destroy the \textsl{conclusive} transfer feature
of both the improved protocol and the original one, because the
measurement result of ``1" may also implies the \textsl{N}th site of
spin chain (1) is in state
$\rho=|\alpha|^2|0\rangle\langle0|+|\beta|^2|1\rangle\langle1|$ with
very small probability. The effect of probabilistic single
excitation is related with $m$, the site on which the excitation
happens. Compared with the probabilistic single excitation, the
effect of probabilistic collective excitations is much more severe,
and exponentially decrease with the length of the channel $N$.

Engineering the modulated spin chain\cite{Mattias} that can
perfectly transfer quantum states is difficult in experiment.
Because the coupling constant between the \textsl{i}th spin and the
\textsl{(i+1)}th spin is $J_i=\sqrt{i*(N-i)}$ \cite{Mattias}, the
ratio of central couplings to the couplings of ends is very large
for long spin chains. To engineer a spin chain with coupling
constants $J_i$ changing in so large range is very difficult in
experiment, and it is hard to maintain the same precision for every
couplings in such spin chains.\vspace{4mm}

\centerline{\textbf{ACKNOWLEDGEMENT}}

We acknowledge all the collaborators of our quantum theory group at
the Institute for Theoretical Physics of our university. This work
was funded by the National Natural Science Foundation of China under
Grant No. 60573008.


\end{document}